\documentclass{ctr_summer}

\usepackage{ctrfont}
\usepackage{natbib}

\usepackage{graphicx}
\usepackage{epstopdf}

\usepackage{amsmath,rotating}

\graphicspath{{figures/}}

\usepackage{url}

\usepackage{color}

\newcommand{\E}{\mathrm{E}}
\newcommand{\mm}{\mathrm{mm}}

\title{Parallel Optimization for LES}

\shorttitle{LES optimization}

\author{C. Talnikar\footnote[1]{Aerospace Computational Design Laboratory, Massachusetts Institute of Technology}, P. Blonigan\footnotemark[1],
J. Bodart\footnote[2]{University of Toulouse ISAE, France}
\and Q. Wang\footnotemark[1]}

\shortauthor{Talnikar et~al.}

\begin{document}

\setcounter{page}{1}

\maketitle
We developed a parallel Bayesian optimization algorithm for large eddy
simulations.  These simulations challenge optimization methods
because they take hours or days to compute, and their objective function
contains noise as turbulent statistics that are averaged over a finite time.
Surrogate based optimization methods, including Bayesian optimization,
have shown promise for noisy and expensive objective functions.
Here we adapt Bayesian optimization to minimize drag in a
turbulent channel flow and to design the trailing edge of a turbine blade to
reduce turbulent heat transfer and pressure loss.  Our optimization
simultaneously runs several simulations, each parallelized to thousands
of cores, in order to utilize additional concurrency offered by today's
supercomputers.
\\
\hrule

\section{Introduction}
In
modeling turbulent flows, large eddy simulation (LES) can reliably capture flow separation and other
phenomena that traditional models including Reynolds averaged Navier-Stokes (RANS) struggle with.
Using LES for design and optimization is not only desirable, but also is made affordable by
increasingly powerful computing clusters. Despite growing computing power, optimization with LES is algorithmically challenging.
This is because of the unsteady, potentially chaotic dynamics of LES, which introduces a sampling error in the
turbulent statistics \citep{oliver2014}. Turbulent statistics are long-time averages of quantities of interest in a turbulent fluid flow. 
\begin{equation}
    \bar{J} = \E[J] = \lim_{T \to \infty} \frac{1}{T} \int_0^T J(u(t)) \ dt
\end{equation}
The infinite time average is approximated by a sample average, which introduces a sampling error in the turbulent statistic.
Because of chaotic
dynamics, totally different sampling errors can come from LES of almost identical designs, starting from the
same initial condition.

For performing optimization with LES we need to consider algorithms that are robust to 
noisy function evaluations. As computing gradients accurately in an LES is expensive we considered
 only derivative free optimiztion techniques \citep{rios2013derivative}. Surrogate-based methods work quite well for such problems as
they filter out the 
noise in the evaluations. In recent years Bayesian optimization is emerging as a promising technique
for optimizing noisy and expensive black box functions \citep{jones1998efficient}. It uses Gaussian processes for fitting a
surrogate to the function evaluations. Successive evaluation points are decided by using a metric
like expected improvement (EI).

A typical LES has two stages; the transient stage
followed by the quasi-steady stage.  Only during the quasi-steady stage are
the statistics sampled. The transient stage, needed merely for reaching
the quasi-steady stage, can be comparable to or
longer than the quasi-steady state.  To avoid wasting time on the transient
stage of a new design, particularly when it is similar to an old design
that has already been simulated into quasi-steady stage, we may opt to
continue simulating the old design. We call this technique snapping.
In addition, the optimization runs in parallel by evaluating multiple
designs at the same time, allowing it to scale
to larger supercomputers. Bayesian optimization using Gaussian processes
provides a good framework for formulating these ideas into an algorithm.

\section{Bayesian optimization}
Bayesian optimization fits function evaluations using a Gaussian process (GP).
A GP is a collection of infinite random variables, defined by a mean function ($m(x)$) 
and a covariance function ($k(x,x^{\prime})$). If $f(x)$ is the true function, then the GP is given by \citep{rasmussen2006gaussian}
\begin{align}
    m(x) &= \E[f(x)], \\
    k(x, x^{\prime}) &= \E[(f(x)-m(x))(f(x^{\prime})-m(x^{\prime}))],
\end{align}
One of the simplest choices for the covariance function is the squared exponential kernel ($k(x,x^{\prime}) = e^{-(\frac{|x-x^{\prime}|}{c_l})^2}$). 
All realizations of the GP are smooth, infinitely differentiable, and
vary over the length scale $c_l$.

A GP can explicitly model noise in the function evaluations.  The noise in an
LES is the sampling error, the
variance of which can be estimated \citep{oliver2014}.  During
optimization, some designs have noisier objective functions than other
designs.  GPs provide a way to incorporate this information in the fitting process. Consider a set of sample points $x_{*}$ having
the function evaluations $y_{*}$, with variance $\sigma_*^2$. The evaluations $y, \sigma^2$ at $x$ can be found from the GP using the following formula
\begin{align}
    y(x) &= k^T_*(K + \sigma_*^2I)^{-1}y_*, \\
    \sigma^2(x) &= k(x_*, x_*) - k^T_*(K + \sigma_*^2I)^{-1}k_*,
\end{align}
where $K = k(x, x), k_* = k(x,x_*)$.  The hyperparameters including the
characteristic length $c_l$ can be decided using a Bayesian model selection.

Once the GP is constructed, we must decide what designs to
simulate next.  This decision must fulfil two competing goals:
exploration and exploitation.
Exploration means the objective should be evaluated in regions where the uncertainty(noise) is high to improve the quality of the surrogate. Exploitation
means the objective should be evaluated at the minimum of the surrogate to get the precise value of the optimum. A metric that provides a good balance
between the two is expected improvement (EI). It is defined by the following formula
\begin{align}
    \mathrm{EI}(x) = \E[\max(f_{min}-Y(x), 0)],
\end{align}
where $f_{min}$ is the current best value of the objective and $Y$ is the random variable corresponding to point $x$. EI has a compact analytical form by computing the
expectation as an integral \citep{snoek2012practical}
\begin{align}
    \mathrm{EI}(x) =
    (f_{min}-y(x))\Phi\left(\frac{f_{min}-y(x)}{\sigma(x)}\right) +
    \sigma(x)\phi\left(\frac{f_{min}-y(x)}{\sigma(x)}\right),
\end{align}
where $\phi$ is the standard normal density and $\Phi$ is the distribution function. The 
point in the design space, which maximizes EI, is chosen as the next
evaluation point.

EI works well for uniformly noisy objective functions, but as we know in an LES the objectives have heterogeneous noise, hence EI needs to
be adapted for such cases. A recent proposal is to multiply EI by a penalization function to discount points that have already been evaluated. It is known 
as augmented expected improvement (AEI) and the multiplier is given by \citep{picheny2013quantile}
\begin{align}
    M(x) = 1 - \frac{\tau}{\sqrt{\sigma^2(x) + \tau^2}},
\end{align}
where $\tau$ is a tunable parameter and $\sigma$ is the standard deviation at a point $x$ evaluated from the GP.
A more rigorous formulation is the expected quantile improvement (EQI). Because the evaluations
are noisy, choosing $f_{min}$ in EI to be the current best evaluation
can be misleading. EQI instead defines improvement on the basis of minimum $\beta$-quantile $q_n(x) = y(x) + \Phi^{-1}(\beta)\sigma(x)$. EQI is defined as
\begin{align}
    \mathrm{EQI}(x) = \E[\max(\min(q_n(x)) - q_{n+1}(x), 0)],
\end{align}
EQI can also be computed analytically and the expression turns out to be quite similar to EI, except that the mean and the variance are conditional.

Today's computers are massively parallel, with more computing cores than
the strong scaling limit of many simulations.  To utilize additional
concurency, we want our optimization to simulate multiple designs in parallel.
Bayesian optimization using EI, however, is inherently serial. At each step EI is maximized, a single point is evaluated and the process is repeated. 
The EI criterion needs to be modified to support maximization over multiple points, so that multiple points can be evaluated at the same time
in parallel. A natural extension is the following \citep{ginsbourger2008multi}
\begin{align}
    \mathrm{EI}(x_1, x_2, ... , x_n) = \E[\max(f_{min} - \min(Y(x_1), Y(x_2) ..., Y(x_n)), 0)],
\end{align}
A problem with the above expression is that it cannot be computed analytically for $n > 2$ and requires expensive Monte Carlo evaluations.
The practical approach is to use an approximation. To avoid computing joint distributions needed for multi-point EI, a sequential
multi-point EI can be performed. The basic idea is to successively do 1-point EIs by conditioning them on the point computed in the previous step.
This still requires a value for the objective at the new points; a safe strategy is to simply use the mean of the GP at that point. This is known as the
Kriging believer strategy and the main problem is that the optimizer may get trapped in non optimal regions if the initial GP fit is bad. An alternative
is to use the constant liar strategy in which a constant value ($L$) is used as the objective evaluation at all the points computed in the multi-point EI. Choices
for L are $\min(y), \max(y)$. Higher $L$ leads to more explorative optimizers.

Apart from exploration and exploitation, in LES it is also
important to consider whether to continue a previous evaluation. If instead of 
simulating a new design, a simulation at an old design is extended we say that 
the new design has snapped on to the old design. This will reduce the
uncertainty at the design point and can improve the quality of the GP fit. 
It is also much cheaper to extend an evaluation than to start a new one because of 
the large transient times in a turbulent fluid flow. There are many possible criteria that can be used to decide when to snap to a previous evaluation or
start a new one. One idea is to check if the two evaluation points are close by. If the Euclidean distance between the two points is below a certain threshold 
then the new point is snapped onto the previous one. A problem with this method is that it cannot be non-dimensionalized easily and a certain amount of tuning 
might be required for each case. Also in higher dimensions the probability of snapping drastically reduces as the design space is much larger. An alternative
is to check if the points are close by using the relative difference between EI at the two points. There is still a certain amount of tuning to be done, but this
idea follows the principle of evaluating points that lead to an improvement in the optimal and surrogate fit.

Before starting the optimization it is important to have a good set of
evaluations from which the initial surrogate is created. For the design of experiment 
Latin hypercube sampling was used to find the design points for the initial evaluations. 

\section{Turbulent channel drag reduction}

\begin{figure}
\centering
\includegraphics[width=0.5\textwidth,natwidth=595,natheight=842]{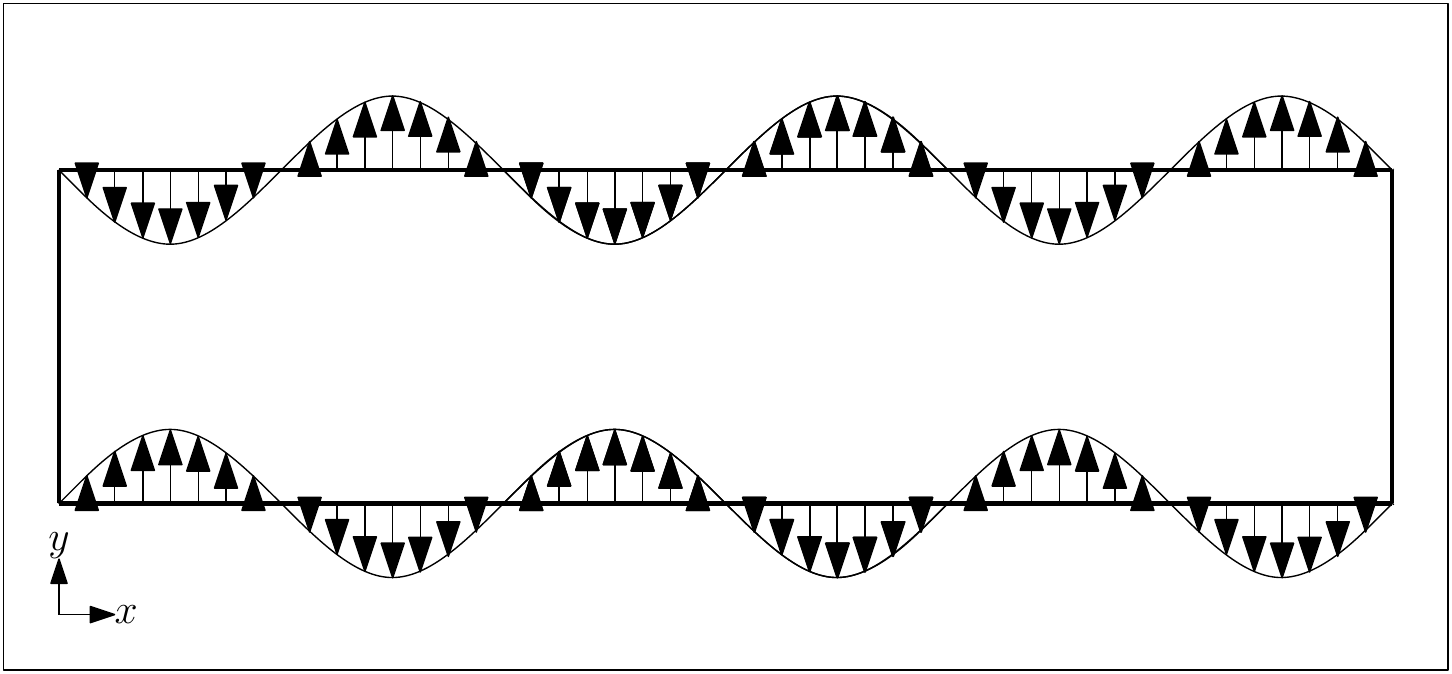}
\caption{Diagram of the traveling wave boundary conditions used for
drag reduction. The $z$ direction points out of the page. }
\label{f:3D_channel_scheme2D}
\end{figure}

To demonstrate our parallel Bayesian optimization algorithm, we consider
flow control for drag reduction of a turbulent channel. Specifically, we
consider the traveling waves studied by \cite{min2006}, \cite{moarref2010controlling}, and \cite{lieu2010controlling}. By enforcing
sinusoidal inflow/outflow at the walls as shown in Figure~\ref{f:3D_channel_scheme2D}, the Reynold's shear stresses can be
modified in the near wall region, resulting in changes to the drag of
the channel. It has been shown that for certain amplitudes, if these
waves are made to propagate upstream, the channel drag will be smaller
than that for laminar flow-through the same channel with the same mass
flow rate (\cite{min2006}). 

We have used our optimization algorithm to find the wave speed
$c$ for which the most drag reduction is achieved for a given mass flow
rate. To this end, we used a time-averaged fractional change in drag as an
objective function

\[
J = \frac{\Delta D}{D_{baseline}} = \frac{D - D_{baseline}}{D_{baseline}},
\]

\noindent where $D$ is the time-averaged drag of the channel and $D_{baseline}$ is
the time-averaged drag of the same channel with no flow control. This
makes the channel optimization a one-dimensional, single-objective
optimization problem.  

Simulations were conducted with the compressible LES solver
Charles$^X$. The mass flow was
kept constant with an additional momentum forcing term. This term was
chosen so that the bulk velocity is 0.2 units. This corresponds to a
Reynolds number of $Re = 13800$ with channel height as the length scale, 
much larger than that used by \cite{min2006}. 


The channel geometry used is 6 $\times$ 2 $\times$ 3 units, with periodic
boundary conditions in the $x$ and $z$ directions. At the walls, the $x$
and $z$ velocities are fixed, while the $y$ velocity had the following
form

\[
v(x) = \pm a \cos\left(\frac{\pi}{3}(x - ct) \right),
\]

\noindent where $a$, the wave amplitude is positive on the lower wall and negative
on the upper wall. For our optimization, $a$ was fixed at $0.04$ units,
or 20\% of the channel bulk velocity. 

A 128 $\times$ 128 $\times$ 128 structured mesh of quads is used, where the cell
sizes are uniform in the $x$ and $z$ directions. In the $y$ direction, the
cell nearest to the wall has an edge length of 0.0015 units. 

The simulations were run for roughly 200 time units for each function
evaluation. This includes 100 time units, or 20 flow-through times, of
transient data that is ignored when computing the objective function 
$\Delta D/D_{baseline}$. The objective function is
computed from the momentum source term used to drive the channel flow,
which is available as an output from Charles$^X$. 

Two key features of our algorithm, parallelism and snapping were
demonstrated in finding the optimal wave speed for drag reduction in a
turbulent channel. For each iteration of the optimization algorithm, two
function evaluations were done in parallel. Snapping was enabled but
only used twice. 

\begin{figure}
\begin{center}
\includegraphics[width=0.7\textwidth]{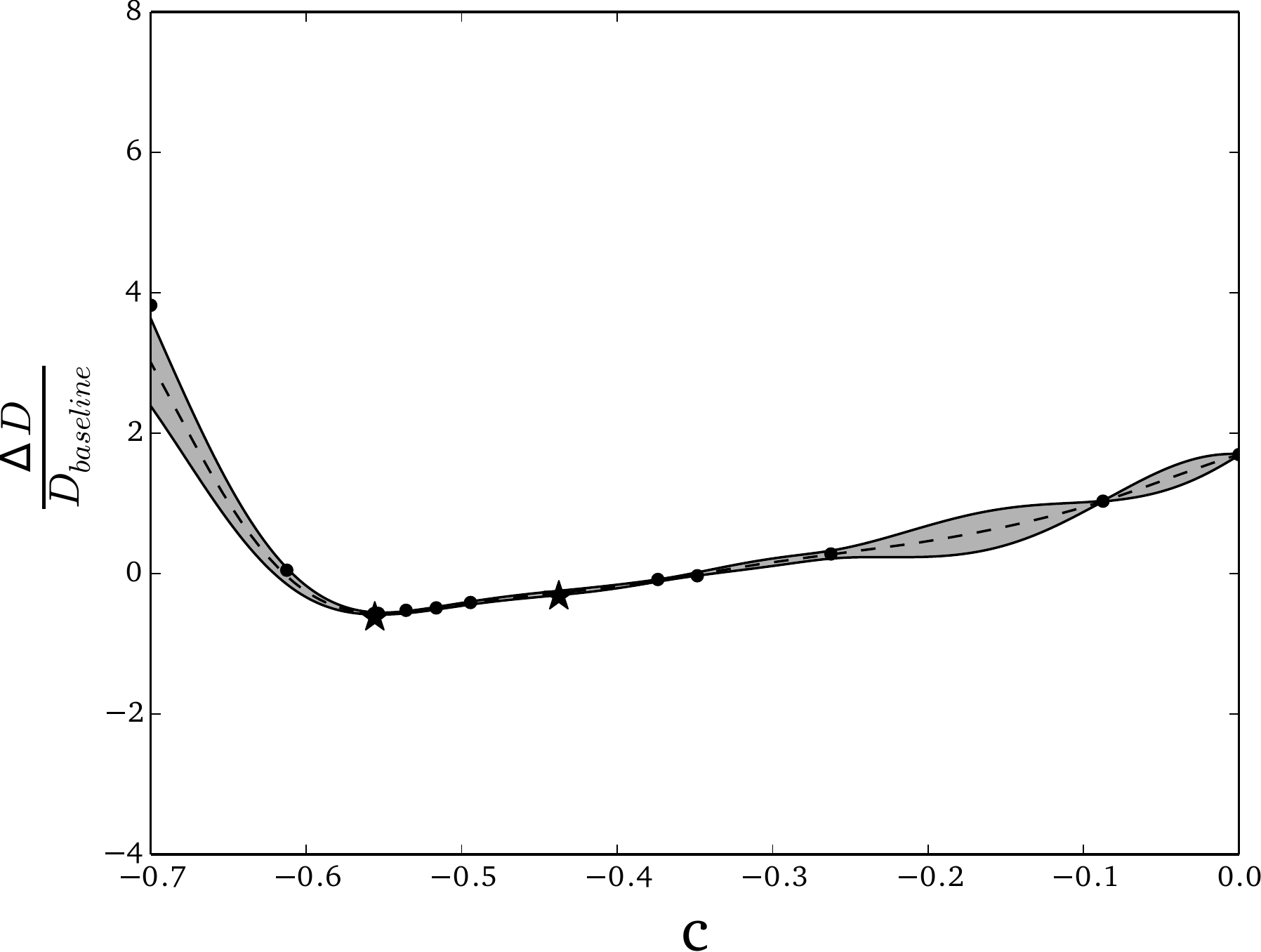}
\caption{GP fit for the change in drag from the baseline (no-control) case $\Delta
D/D_{baseline}$ versus
wave speed $c$. Negative values of $c$ correspond to upstream traveling
waves. The region bounded by one standard deviation $\sigma$ of the GP
is shaded grey. (\dashed) GP mean,(\solidcircle) evaluations, (\starsol) snapped evaluations. 
\label{f:channel_GP}
}
\end{center}
\end{figure}

The optimization results are shown in Figure~\ref{f:channel_GP}. The
optimal wave speed was found to be $c = -0.5703$ (the wave speed $c$ is non-dimensionalized as wave velocity divided by
the speed of sound), resulting in around
60\% drag reduction from the baseline case with no traveling wave. Note the large number
of function evaluations in the vicinity of the minimum, including the
two points where snapping was used at $c=-0.4375$ and $c=-0.5590$. This
shows that the decisions made using EI result in an effective use of
computational resources. 

The trend in fractional drag change is consistent with that found by
\cite{min2006} for low-magnitude values of wave speed $c$. As wave speed
increases in magnitude upstream, the drag decreases through the changes in
the Reynolds stresses near the wall. Eventually, the Reynolds shear
stress $\bar{u'v'}$ becomes negative, and the drag falls to sub-laminar 
values near the optimum (drag of a laminar channel with no
control is 55\% smaller than that of a turbulent channel with no
control \cite{min2006}). The increase in drag to the left of the
optimum was not observed in literature, and arises from
compressibility effects. At $c=-0.6$ and less, acoustic waves dominate
the flow field, which is not surprising considering the nearly transonic
speeds attained by the traveling waves.

\begin{figure}
\begin{center}
\includegraphics[width=0.48\textwidth,trim=5cm 0.5cm 5cm 3cm,clip]{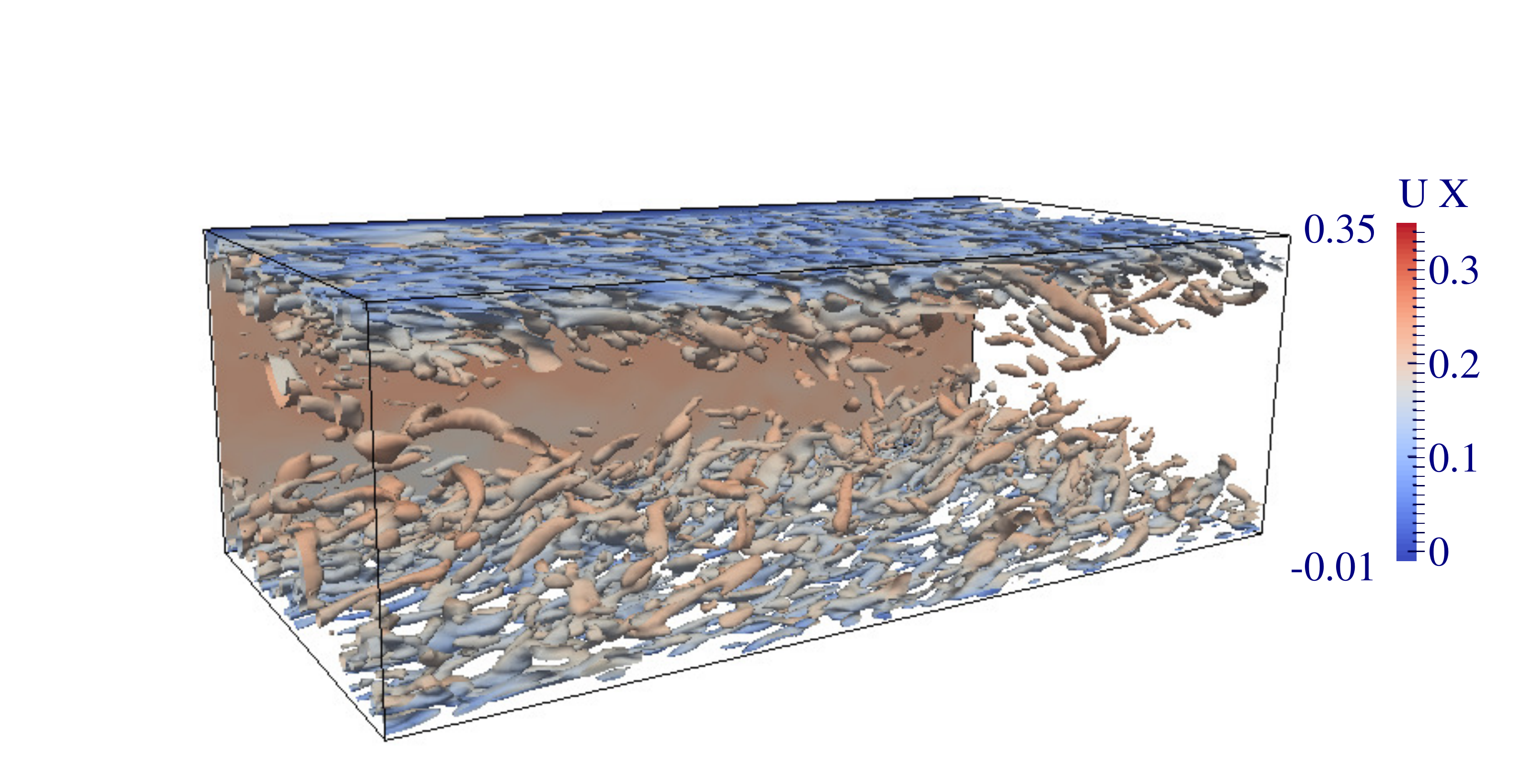}
\includegraphics[width=0.48\textwidth,trim=5cm 0.5cm 5cm 3cm,clip]{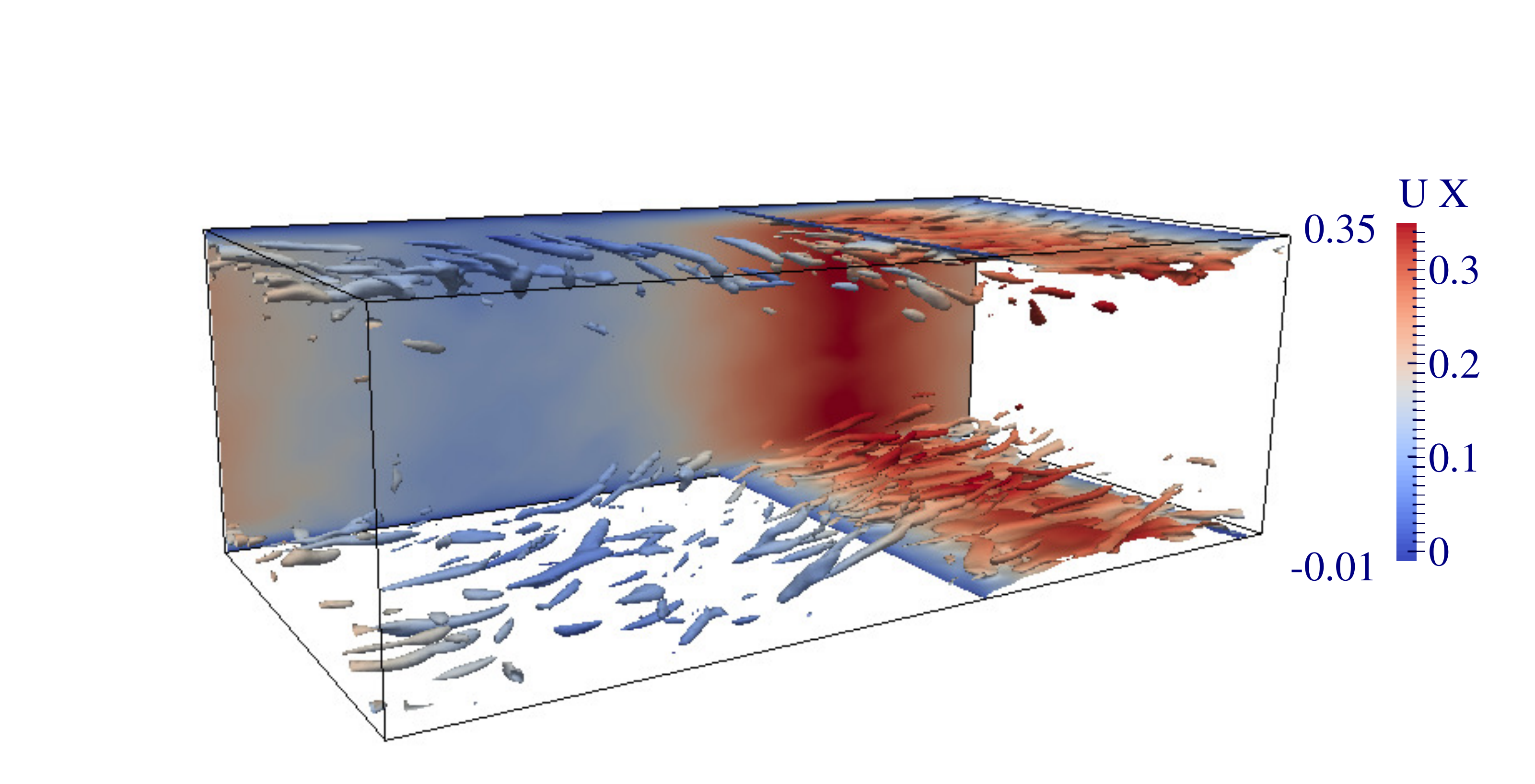}
\put(-300,0){(a)}
\put(-50,0){(b)}
\caption{Q criterion isocontours colored by streamwise-velocity for uncontrolled
(a) and optimally controlled (b) channel flow. The contours
correspond to $Q=0.04$ in both plots.
\label{f:Qcrit_channel}
}
\end{center}
\end{figure}

Finally, we consider the vortical flow structures shown in Figure~\ref{f:Qcrit_channel} for the baseline and optimized cases. Comparing
these plots, we see very few structures in the optimized change relative
to the baseline case. In modifying the Reynolds stresses near the walls,
the traveling waves have destroyed a large number of flow structures. 

Also, it is important to note the sheets on the channel walls. These
sheets correspond to the inflow and outflow boundary conditions: the
sheet begins just downstream of the maximum inflow velocity and ends
near the maximum outflow velocity. One can think of the boundary
conditions as a vortex sheet, where one period of the sinusoid, inflow
followed by outflow just downstream, represents one vortex. The
Q-criterion reveals the presence of these vortices at the walls with the
sheets. 

\section{Turbine blade shape design}
The second test case for the Bayesian optimization formulation is the trailing edge design
of a turbine blade for minimizing the heat transfer and pressure loss.
The blade is a turbine nozzle guide vane designed by researchers at Von Karman Institute (VKI) \citep{arts1992aero}. 
The baseline case for this optimization is the blade design given in the aforementioned paper.
The chord length of the blade is $67.647 \mm$ and the blades are in a linear cascade with the pitch being
$0.85$ times the chord length. The inlet in the simulation setup of the problem is  $100 \mm$ 
upstream of the leading edge of the blade. The inlet flow is at an angle of $55$ degrees to the chord of the
blade.  A linear cascade is simulated by having
periodic boundary conditions at the top and bottom. The spanwise extent of the setup is $10 \mm$. 
The blade surface on the pressure and suction side is assumed to be isothermal, the inlet isentropic Mach
number is 0.9 and the isentropic Reynolds number downstream is $10^6$. The turbulent intensity at the
inlet is $4\%$ and the length scale is $3 \mm$. The compressible flow solver used is CharLES which is 
an explicit finite volume code. Perturbations to the mean flow at the
inlet are injected using a synthetic turbulence generator.

The flow accelerates as it goes around the suction side and reaches a peak of Mach $1$. At the
specified turbulent intensity the boundary layer transitions from a laminar to turbulent at
the suction side about $60 \mm$ along the curved surface starting from the leading edge. The
flow separates at the trailing edge. The shape of the trailing edge can greatly influence the 
seperation locations on the pressure and suction side as well as flow in the
recirculation region. This affects the heat transfer on the surface
of the blade and the pressure loss in the flow. The objective of the optimization is set to be 
a linear combination of the non-dimensional versions of the two. The heat transfer is characterized by
the Nusselt number
\begin{equation}
    Nu = \frac{\bar{h}L}{k},\put(-300,3){(a)}
\put(0,3){(b)}
\end{equation}
where $k$ is the thermal conductivity at $T = 300K$, $0.0454 \frac{W}{mK}$. $L$ is the
trailing edge radius in the baseline case, $0.71\mm$. $\bar{h}$ is the average 
heat transfer coefficient integrated over time and
the part of the blade from $28\mm$ downstream of the leading edge up to the trailing
edge. The formula for $\bar{h}$ is 
\begin{equation}
    \bar{h} = \frac{1}{S(t_j-t_i)\Delta T}\int_{t_i}^{t_j} \int_S k \frac{\partial T}{\partial n} dS dt,
\end{equation}
where $t_i$ is the transient time and $t_j$ is the stop time of the simulation. $S$ is the
surface area. $\Delta T = 120 K$ is the temperature difference between the surface of the blade and the
stagnation temperature of the flow.

The pressure loss is characterized by the pressure loss coefficient.
It is computed as the pressure loss divided by the inlet total pressure $\bar{p}_l = \bar{p}_{t,l}/p_{t,in}$. 
The pressure loss is computed $16 \mm$ downstream of the trailing edge by cutting a plane normal to the inlet flow. The formula for $\bar{p}_{t,l}$ is
\begin{equation}
    \bar{p}_{t,l} = \frac{1}{S(t_j-t_i)}\int_{t_i}^{t_j} \int_S (p_{t,in}-p_{t,p})dS dt
\end{equation}
where $p_{t, in}$ is the inlet total pressure and $p_{t,p}$ is the total pressure at the 
plane given by $p_p (1 + \frac{\gamma-1}{2}M_p^2)^{\frac{\gamma}{\gamma-1}}$, $M_p$ is the Mach number and $p_p$ is the
static pressure at the plane. The time integration for both these quantities is started after an initial transient
time. For this simulation it was chosen to be $0.3$ times the time it takes for a single flow-through based
on the amount of time it takes for the objective to stabilize. Time averaging was then 
performed over $0.7$ flow-through time. The weights in the linear 
combination of the two objective functions were chosen such that both
were of similar magnitude for the baseline case and less than one.

\begin{figure}
\begin{center}
\begin{minipage}[t]{0.4\columnwidth}
\includegraphics[width=1.0\textwidth]{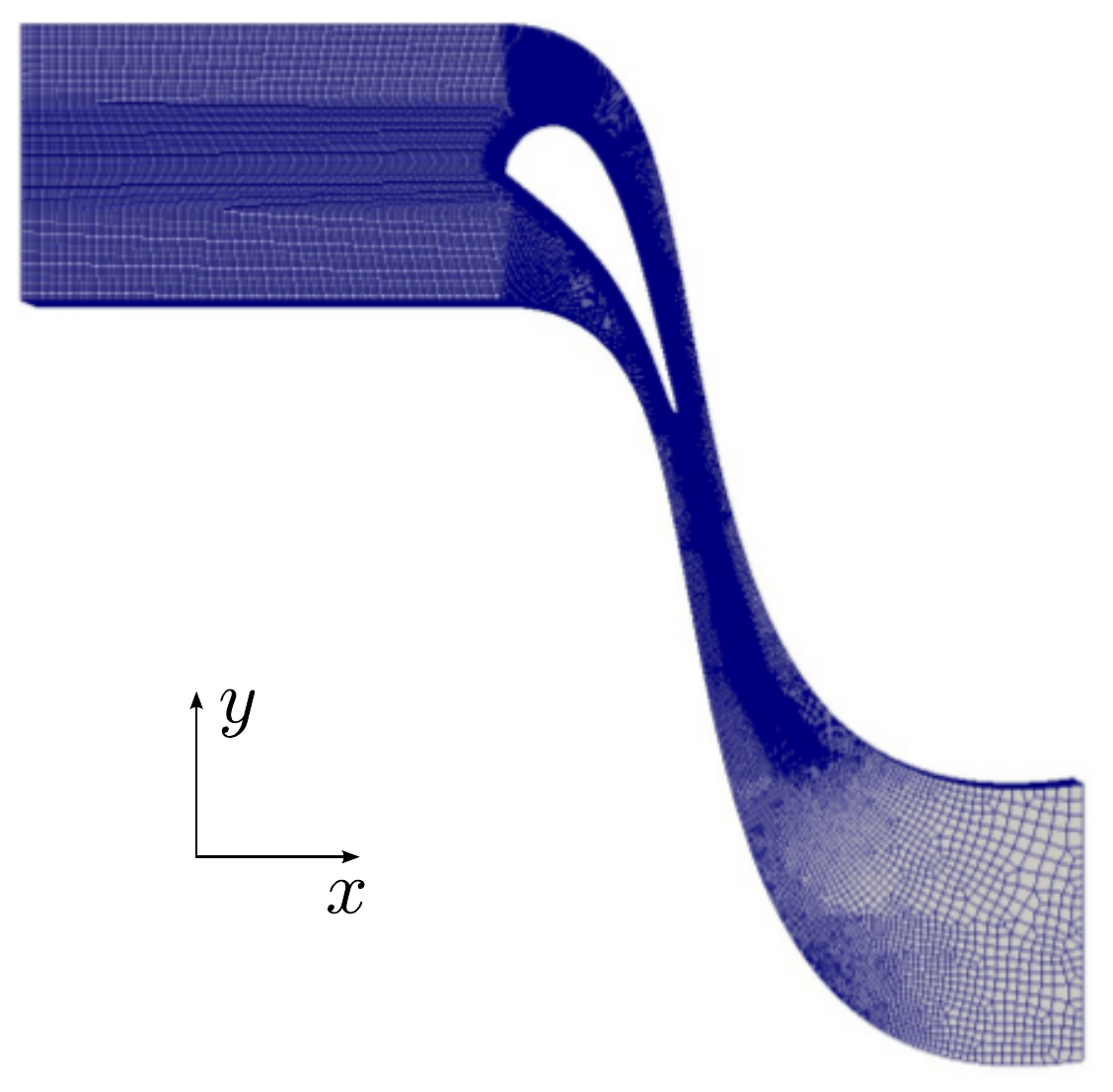}
\label{f:mesh}
\end{minipage}
\hfill
\begin{minipage}[t]{0.5\columnwidth}
\includegraphics[width=0.95\textwidth]{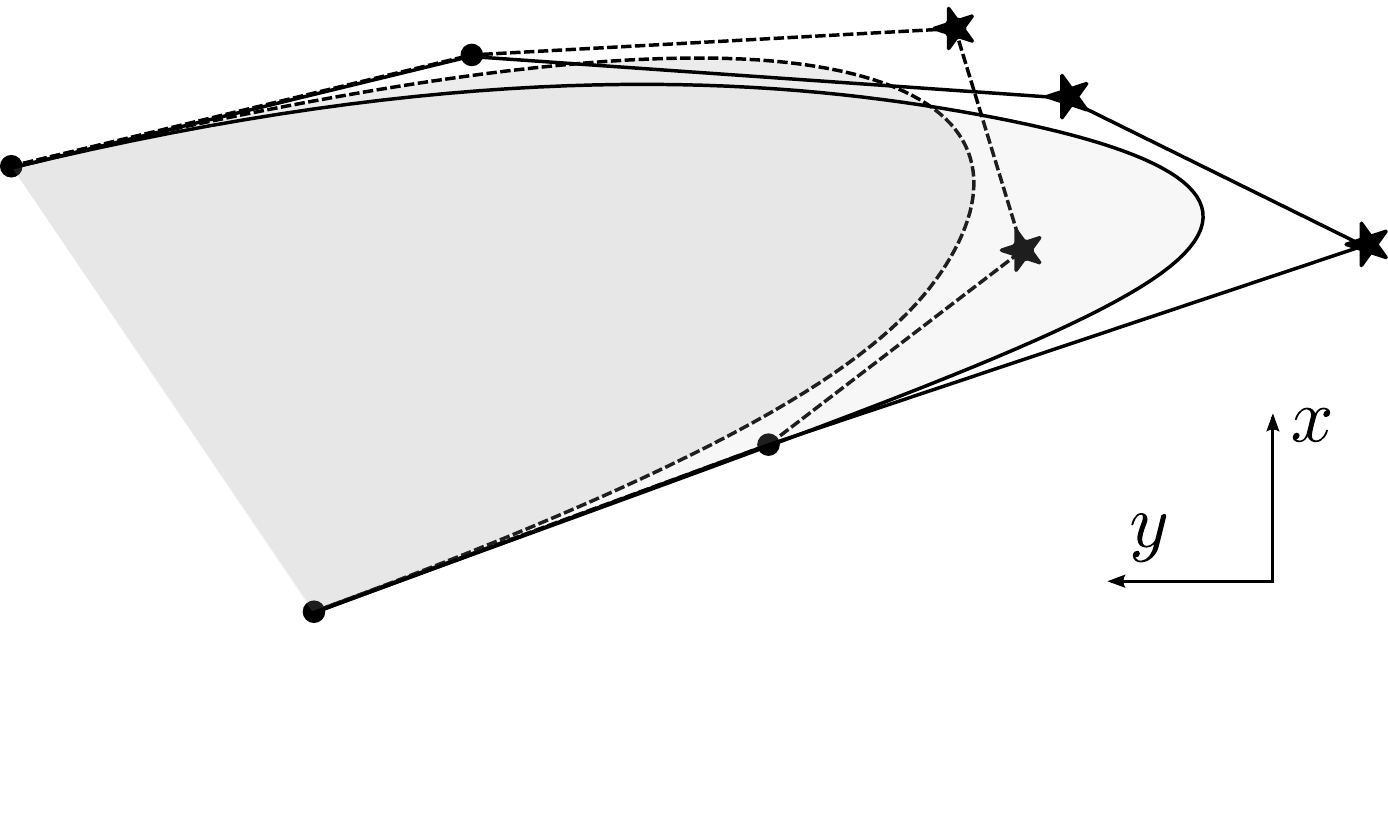}
\put(-300,3){(a)}
\put(0,3){(b)}

\end{minipage}
\caption{(a) Computational mesh around the baseline geometry. (b) Baseline (\dashed) and optimal(\solid) trailing edges. (\solidcircle) indicates fixed points and (\starsol) control points used in the design.
\label{f:trailing_edge_geom}
}
\end{center}
\end{figure}

The mesh for the simulation is a hybrid structured and unstructured mesh as shown in Figure~\ref{f:trailing_edge_geom}(a). 
 The smallest cell size
away from the walls is $0.5 \mm$, which is about a factor of $6$ smaller than the most significant eddies.
The first cell size at the wall is set to $0.02 \mm$, which results in a maximum $y_+$ of 10 over the surface
of the blade to reduce the cost of the simulation \citep{collado2012effects}. 
This means that the mesh for the boundary layer is under-resolved, and a wall model is required
for reliably capturing the heat transfer characteristics at the blade surface. A wall model 
for CharLES which works for blades having a high flow incidence angle, is currently under development.
In the mean time, the simulation was run without a wall model to validate the performance of our optimizer.

To parameterize the trailing edge in two dimensions we used B-splines. They are piecewise polynomial functions of a specific 
order that can be used to represent curves. To limit the number of parameters needed to characterize the
curve, the order of the B-spline was chosen to be $3$ with $2$ control
points as shown in Figure~\ref{f:trailing_edge_geom}(b). By moving these points
we can change the shape of the B-spline. Two control points translate to $4$ degress of freedom. The
range of the parameters (locations of the $2$ control points) was chosen such that the trailing 
edge is not too sharp and there is no loop present in the curve. For the actual meshing, a set
of points defining the entire blade (fixed part + trailing edge curve) is passed to the ANSYS ICEM meshing tool that generates the mesh. 
The mesh topology is assumed to be constant as
the changes to the shape of the blade are local and restricted to the trailing edge.

The choices of objective function and design parameters make the blade
optimization a four-dimensional, single-objective optimization
problem. In this optimization, 10 points were evaluated in the design of experiment and 25 points were evaluated using
the parallel EQI criterion. The number of points evaluated in parallel were 4. 
The value of the objective in the baseline case is 0.6877 (Nusselt number: 6604, Pressure loss coefficient: 0.02241) with a 
standard deviation of 0.0140. The value of the objective for the optimal case (Figure~\ref{f:trailing_edge_geom}(b)) is 
0.5587 (Nusselt number: 5486, pressure loss coefficient: 0.01775). 
We get a 17\% reduction in heat transfer and 
a 21\% reduction in pressure loss. During the optimization
run, the criterion twice snapped onto the final design, i.e., it chose to
continue simulating this design instead of starting to simulate a similar
design proposed by EQI.

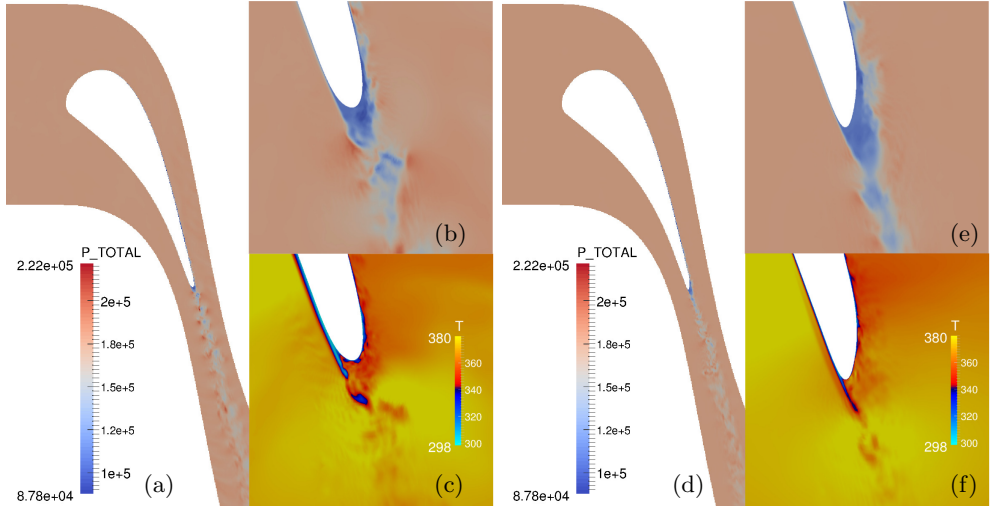
\begin{figure}
\begin{center}
\begin{minipage}[t]{0.48\columnwidth}
\centering
\includegraphics[width=1.0\textwidth]{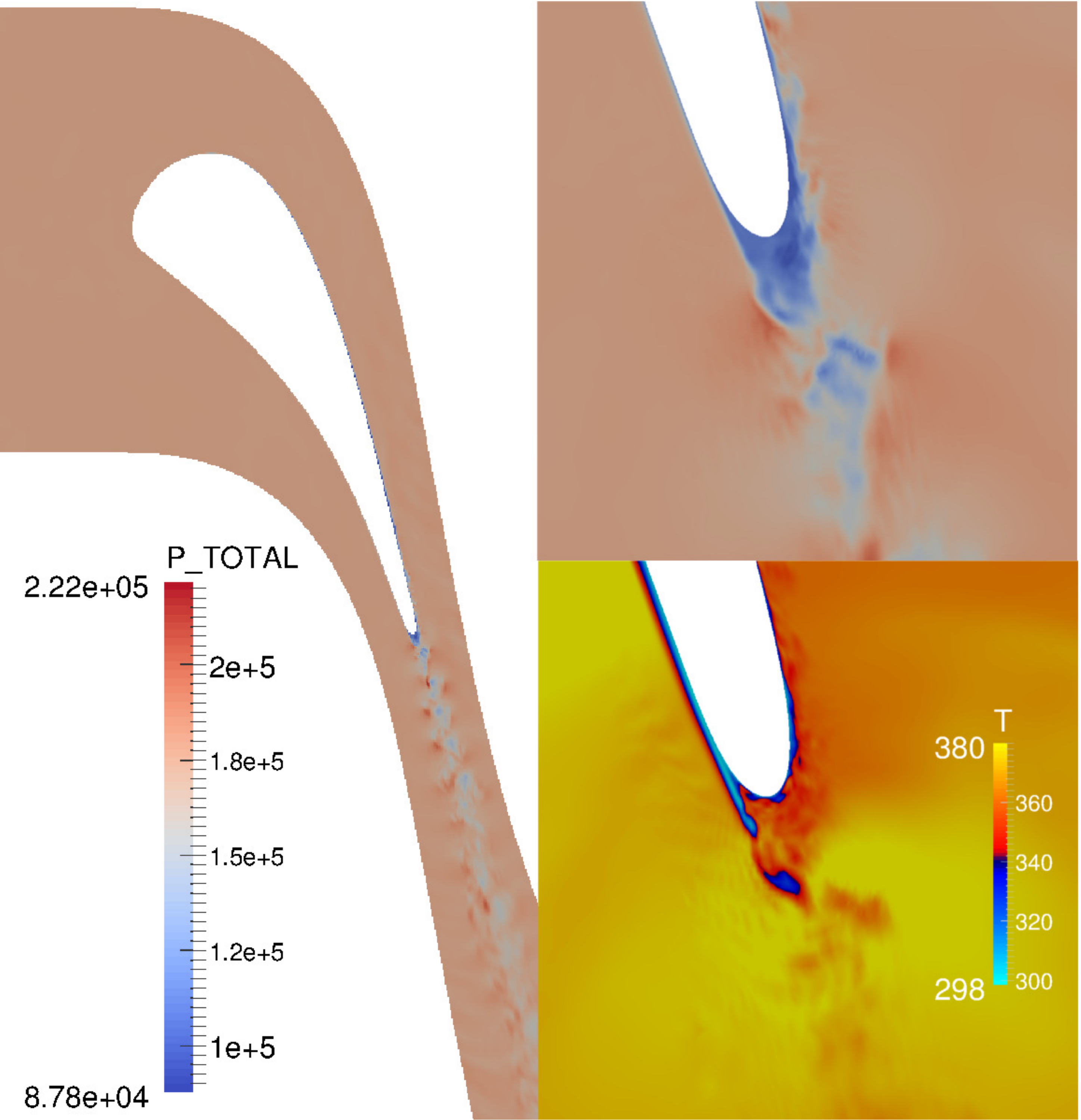}
\end{minipage}
\begin{minipage}[t]{0.48\columnwidth}
\centering
\includegraphics[width=1.0\textwidth]{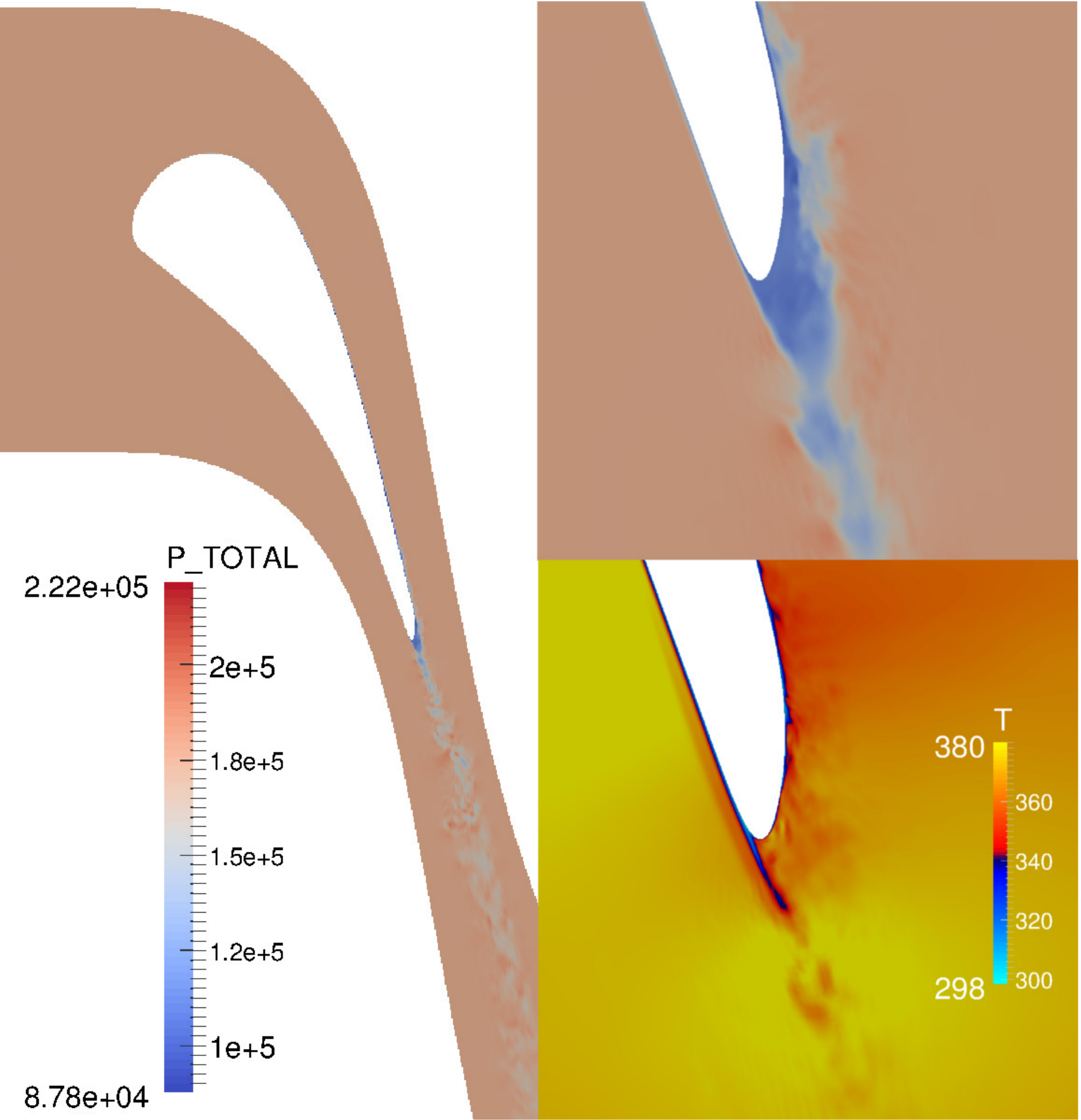}
\end{minipage}
\put(-320,5){(a)}
\put(-210,100){(b)}
\put(-210,5){(c)}
\put(-120,5){(d)}
\put(-15,100){(e)}
\put(-15,5){(f)}
\caption{(a) Baseline geometry with total pressure contours in Pascal (b) Zoom in the trailing edge region. (c) Map of the temperature in Kelvin.
(d-e-f) Same plots for the optimal geometry.
\label{f:blade_flow}
}
\end{center}
\end{figure}

In the Figure~\ref{f:blade_flow}, we compare the flow fields near the trailing edge of the
baseline and optimal case. From Figure~\ref{f:blade_flow}(e), we see that the optimal design
is skewed toward the pressure side. This design ensures that separation
on the pressure side is considerably downstream of the separation on the
suction side, unlike the baseline case. 

The optimizer moves the separation point downstream on the pressure side 
because the boundary layer is laminar. Because of this, the mixing layer between the
recirculation region and the fluid on the pressure side is much thinner
than that on the suction side. This allows the hot fluid outside the
pressure side boundary layer to enter the recirculation region with
relative ease, as is evident in the temperature contour shown in Figure~\ref{f:blade_flow}(f). Once this fluid enters the recirculation region, it flows
back towards the trailing edge, resulting in the higher heat fluxes at
the trailing edge. By moving the pressure side separation point downstream,
the mixing layer is moved downstream, and as a result the portion of the mixing
layer adjacent to the recirculation region is more stable. This results
in less mixing and less hot fluid in the recirculation region, which in
turn reduces the heat transfer to the trailing edge. 

At the same time, the separation point on the suction side is moved
upstream, and a larger portion of the trailing edge is exposed to the
recirculation region. However, because the boundary layer on the suction
side is turbulent, the fluid in this region is well mixed. Because the
fluid near the wall is much cooler than the fluid outside the boundary
layer, the fluid in the turbulent boundary layer and the portion of the
recirculation region adjacent to the suction side of the blade is
relatively cool. 

Additionally, the spacing between the separation
regions on the pressure and suction side affects the wake structure
considerably. In Figure~\ref{f:blade_flow}(a), we can see coherent structures in the total pressure 
contours indicating vortices shed from the trailing edge for the
baseline case. However, these distinct structures are absent from the
optimized design. The absence of these vortex structures implies less
large-scale mixing, which is consistent with the considerable reduction
in pressure loss that is achieved by the optimized design.

\section{Conclusion}

In summary, we have demonstrated a novel parallel Bayesian optimization
framework
for flow control for a turbulent channel flow and for the design of the
trailing edge of a turbine inlet guide vane. The framework uses a parallel expected quantile improvement (EQI)
criterion to determine whether to explore new designs or to exploit
existing designs to reduce the local error in design space by continuing existing LES.
Furthermore, it
can explore designs in parallel, allowing the use of as many
computational resources as desired. 

In both test cases our framework found a vastly improved design from the
baseline case. To find these optimized designs, the framework employed
novel features such as snapping and a parallel function evaluations. 

Given the promising results of our parallel Bayesian optimization
framework, we plan to apply it to the turbine blade case presented
above with a working wall model. Additionally, we plan to apply our
framework to other applications, to demonstrate its utility to 
wider range of engineers and scientists.

\subsection*{Acknowledgements}
We thank Jasper Snoek and Ryan Adams from Harvard University
for sharing their expert knowledge on Bayesian optimization.
Mihailo Jovanovic and Armin Zare from University of
Minnesota have contributed to optimization of the travelling wave in
turbulent channel.  The turbine trailing edge optimization problem is
motivated by Gregory Laskowski and Michael
Hayek from GE Aviation and Sriram Shankaran from GE Global Research.
Sanjeeb Bose and Sanjiva Lele from Stanford University and Gaofeng Wang
from Zhejiang University helped formulate
the trailing edge optimization problem.
The first and last authors acknowledge funding from GE Aviation and
GE Global Research.
The LES are performed on Cetus/Mira supercomputers at
Argonne National Lab, using the CharLES and CharlesX solver.
We particularly thank the CTR Summer Program for
making possible this massively collaborative work.


\bibliographystyle{ctr}
\bibliography{ctr_summer_lesopt}

\begin{thebibliography}{12}
\expandafter\ifx\csname natexlab\endcsname\relax\def\natexlab#1{#1}\fi

\bibitem[Arts \& de~Rouvroit(1992)]{arts1992aero}
{\sc Arts, T. \& de~Rouvroit, M.~L.} 1992 Aero-thermal performance of a
  two-dimensional highly loaded transonic turbine nozzle guide vane: A test
  case for inviscid and viscous flow computations. {\em J. Turbomach.\/} {\bf
  114}~(1), 147--154.

\bibitem[Collado~Morata {\em et~al.\/}(2012)Collado~Morata, Gourdain, Duchaine
  \& Gicquel]{collado2012effects}
{\sc Collado~Morata, E., Gourdain, N., Duchaine, F. \& Gicquel, L.} 2012
  Effects of free-stream turbulence on high pressure turbine blade heat
  transfer predicted by structured and unstructured les. {\em Int. J. Heat.
  Mass. Tran.\/} {\bf 55}~(21), 5754--5768.

\bibitem[Ginsbourger {\em et~al.\/}(2009)Ginsbourger, Le~Riche, Carraro {\em
  et~al.\/}]{ginsbourger2008multi}
{\sc Ginsbourger, D., Le~Riche, R., Carraro, L. {\em et~al.\/}} 2009 A
  multi-points criterion for deterministic parallel global optimization based
  on gaussian processes. {\em J. Global Optim., in revision\/} .

\bibitem[Jones {\em et~al.\/}(1998)Jones, Schonlau \&
  Welch]{jones1998efficient}
{\sc Jones, D.~R., Schonlau, M. \& Welch, W.~J.} 1998 Efficient global
  optimization of expensive black-box functions. {\em J. Global Optim.\/} {\bf
  13}~(4), 455--492.

\bibitem[Lieu {\em et~al.\/}(2010)Lieu, Moarref \&
  Jovanovi{\'c}]{lieu2010controlling}
{\sc Lieu, B.~K., Moarref, R. \& Jovanovi{\'c}, M.~R.} 2010 Controlling the
  onset of turbulence by streamwise travelling waves. part 2. direct numerical
  simulation. {\em J. Fluid Mech.\/} {\bf 663}, 100--119.

\bibitem[Min {\em et~al.\/}(2006)Min, Kang, Speyer \& Kim]{min2006}
{\sc Min, T., Kang, S., Speyer, J. \& Kim, J.} 2006 Sustained sub-laminar drag
  in a fully developed channel flow. {\em J. Fluid Mech.\/} {\bf 558},
  309--318.

\bibitem[Moarref \& Jovanovi{\'c}(2010)]{moarref2010controlling}
{\sc Moarref, R. \& Jovanovi{\'c}, M.~R.} 2010 Controlling the onset of
  turbulence by streamwise travelling waves. part 1. receptivity analysis. {\em
  J. Fluid Mech.\/} {\bf 663}, 70--99.

\bibitem[Oliver {\em et~al.\/}(2014)Oliver, Malaya, Ulerich \&
  Moser]{oliver2014}
{\sc Oliver, T., Malaya, N., Ulerich, R. \& Moser, R.} 2014 Estimating
  uncertainties in statistics computed from direct numerical simulation. {\em
  Phys. Fluids\/} {\bf 26}.

\bibitem[Picheny {\em et~al.\/}(2013)Picheny, Ginsbourger, Richet \&
  Caplin]{picheny2013quantile}
{\sc Picheny, V., Ginsbourger, D., Richet, Y. \& Caplin, G.} 2013
  Quantile-based optimization of noisy computer experiments with tunable
  precision. {\em Technometrics\/} {\bf 55}~(1), 2--13.

\bibitem[Rasmussen \& Williams(2006)]{rasmussen2006gaussian}
{\sc Rasmussen, C. \& Williams, C.} 2006 {\em Gaussian Processes for Machine
  Learning\/}. MIT Press.

\bibitem[Rios \& Sahinidis(2013)]{rios2013derivative}
{\sc Rios, L.~M. \& Sahinidis, N.~V.} 2013 Derivative-free optimization: A
  review of algorithms and comparison of software implementations. {\em J.
  Global Optim.\/} {\bf 56}~(3), 1247--1293.

\bibitem[Snoek {\em et~al.\/}(2012)Snoek, Larochelle \&
  Adams]{snoek2012practical}
{\sc Snoek, J., Larochelle, H. \& Adams, R.~P.} 2012 Practical bayesian
  optimization of machine learning algorithms. In {\em Adv. Neur. In.\/}, pp.
  2951--2959.

\end{thebibliography}

\end{document}